\documentclass[aps,prc,reprint,superscriptaddress,nofootinbib]{revtex4-1}
\usepackage{graphicx}
\usepackage{multirow}
\usepackage{comment}
\usepackage{color}

\usepackage{amssymb}
\usepackage{amsmath}
\usepackage{color}
\usepackage{lineno}
\usepackage{scalerel,stackengine}

\stackMath
\newcommand\reallywidehat[1]{%
\savestack{\tmpbox}{\stretchto{%
  \scaleto{%
    \scalerel*[\widthof{\ensuremath{#1}}]{\kern-.6pt\bigwedge\kern-.6pt}%
    {\rule[-\textheight/2]{1ex}{\textheight}}
  }{\textheight}%
}{0.5ex}}%
\stackon[1pt]{#1}{\tmpbox}%
}

\def\nuc#1#2{\relax\ifmmode{}^{#1}{\protect\text{#2}}\else${}^{#1}$#2\fi}

\newcommand{\be}{\begin{eqnarray}}
\newcommand{\ee}{\end{eqnarray}}

\newcommand{\bwt}{\begin{widetext}}
\newcommand{\ewt}{\end{widetext}}

\begin{document}

\title{Inclusive breakup calculations in angular momentum basis: application to $^7$Li+$^{58}$Ni}


\author{Jin Lei}
\email[]{jinl@ohio.edu}

\affiliation{Institute of Nuclear and Particle Physics, and Department of Physics and Astronomy,
Ohio University, Athens, Ohio 45701, USA}

%


\begin{abstract}
The angular momentum basis method is introduced to solve the inclusive breakup
within the model proposed by Ichimura, Austern, and Vincent [Phys. Rev. C 32, 431 (1985)]. 
This method is based on the geometric transformation between Jacobi coordinates,
thus it is easy to corporate with particle spins. 
To test the validity of this partial wave expansion method, a benchmark 
calculation is done comparing with the one given in 
[Phys. Rev. C 92, 044616 (2015)]. Using the distorted-wave Born 
approximation (DWBA) version of IAV model, some applications to $^7$Li reactions 
are presented and compared with available data.
\end{abstract}


\date{\today}%
\maketitle

\section{Introduction \label{sec:intro}}
Investigation of mechanisms responsible for the large inclusive $\alpha$
particle production cross section observed in
breakup of light-weakly bound projectiles (e.g. $^{6,8}$He, $^{6,7}$Li and $^{7,9}$Be) 
is a topic of current interest, both experimentally and theoretically
\cite{Chattopadhyay16,Pandit17,Carnelli18,Sgouros16,Canto15}.
This is a difficult problem, becase different reaction mechanisms, 
like elastic breakup, transfer, compound
nuclear evaporation, inelastic breakup and incomplete fusion contribute
to the $\alpha$ yield.

From the theoretical point of view, one can represent this kind of 
reactions as $a+A \to b+B^*$, where $a=b+x$ and $B^*$ is any possible state
of $x+A$ system. This reaction includes the breakup processes in which $x$
is elastically scatted by $A$ leaving all the fragments in the ground states,
which is usually called elastic breakup (EBU), but also breakup accompanied
by target excitation, particle(s) exchange between $x$ and $A$, $x$ transfer
to $A$, the fusion of $x$ by $A$, which are globally referred to as nonelastic
breakup (NEB). The total breakup (TBU) is therefore the sum of EBU and NEB
components.

The IAV model \cite{IAV85}, which was originally proposed in the 1980s, 
is used to study this inclusive breakup. Due to the computational limitations at that time, this
model was apparently fallen into disuse. Recently , the model model has been
re-examined by several groups \cite{Jin15,Jin15b,Pot15,Pot17,Carlson2016,Moro2016}.
Moreover a systematic study of the alpha productions in $^6$Li induced
reactions has been recently reported by Lei and Moro \cite{Jin17},
in which the numerical calculations using the IAV model agree well
with the experimental data.

For $^7$Li, several experimental groups have reported large alpha yields
and tried to understand the origins of these alphas by
by using Q-value considerations and by direct identification of the 
reaction products \cite{Luong13,Pandit16,Shrivastava13,Pandit17}.
However, a proper interpretation of these alphas
are still lacking. The IAV model, which successfully reproduce the alphas
produced by $^6$Li is a promising tool for this purpose. 
From the theoretical point of view, a important difference between these 
two systems is that the $\alpha+d$ cluster in $^6$Li is in a 
predominantly $\ell=0$ configuration, whereas the $\alpha+t$ cluster 
conforming the $^7$Li system is in a $\ell=1$ configuration. 
This makes the numerical calculation more challenging since
more angular momentum configurations are involved in the calculation.

For this reason, most applications of the IAV formalism have been 
restricted to deuterons and $^6$Li. 
In order to extend the model to other interesting systems, it is 
advisable to test its validity and accuracy for $\ell>0$ cases. 

In the paper, a new method to compute the IAV inclusive breakup 
formula is implemented in a more efficient way. The derived formula has been tested for 
the $\ell=0$ against the previously implemented method. 
This former method becomes numerically difficult for $\ell>0$ cases, 
due to the additional angular momentum couplings 
(details see Appendix B of Ref.\cite{Jin15}). Moreover, the inclusion of the 
intrinsic spins will make the calculation even harder.  
Consequently, an alternative method which can deal with these more 
complicated situations would be advisable.

The paper is organized as follows: In Sec. II we summarize the main formulas 
of the IAV model and outline expansion in angular momentum basis. 
In Sec. III, the formalism is applied to inclusive breakup reactions 
induced by $^7$Li. Finally, in Sec. IV we summarize the main results.

\section{Theoretical Models}
\begin{figure}[tb]
\begin{center}
 {\centering \resizebox*{0.9\columnwidth}{!}{\includegraphics{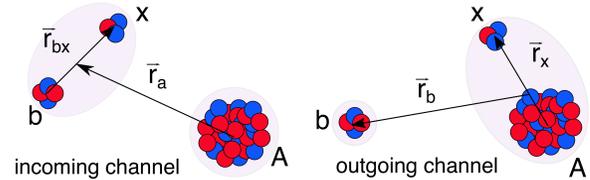}} \par}
\caption{\label{fig:coordinates}(Color online)
Coordinates used in the breakup reaction.}
\end{center}
\end{figure}
In this section, we briefly summarize the model of IAV and introduce
a more effcient method for partial wave expansion comparing with the one used
in Ref.\cite{Jin15}. The new method is more general and easy to incorporate
particle spins.

First, we can write the process under study in the form
\begin{equation}
a(=b+x) + A \to b + B^* ,
\end{equation}
where the projectile $a$, constituted by $b$ and $x$, interacts with the target
$A$, leaving particle $b$ and other fragments. Thus $B^*$ is any possible
state between $x+A$ system.

The effective three body Hamiltonian of this
system is
\begin{equation}
\label{eq:H}
H(\xi)=H_0 + V_{bx} + V_{xA}(\xi) + U_{bA} + H_A(\xi),
\end{equation}
where $H_0$ is the total kinetic energy operator, $V_{bx}$ is the interaction
between the cluster $b$ and $x$, $H_A(\xi)$
is the Hamiltonian of the target nucleus (with $\xi$ denoting its internal coordinates),
and $V_{xA}$ and $U_{bA}$ are fragment-target interactions.

In writing the Hamiltonian of the system in the form (\ref{eq:H}) we make
a clear distinction between the two cluster constituents; the interaction
with the target of the fragment $b$, the one which is assumed to be observed in the experiment,
is described with a (complex) optical potential. Nonelastic processes
arising from this interaction (e.g., target excitation, transfer, 
sequential breakup, and incomplete fusion) are included only effectively through the
imaginary part of $U_{bA}$. Then particle $b$ is said to act as a spectator.
On the other hand, the interaction of the particle $x$ with the target
retains the target degrees of freedom ($\xi$).

By using the closure relation and optical reduction, IAV separated the
inclusive breakup cross section in terms of elastic breakup and nonelastic
breakup, with the latter is given by
\begin{equation}
\label{eq:neb}
\frac{d^2\sigma}{dE_bd\Omega_b}\Big|_{NEB} = -\frac{2}{\hbar \nu_a} \rho_b(E_b)
\langle \psi^0_x(\vec{k}_b)| W_x | \psi^0_x(\vec{k}_b) \rangle,
\end{equation}
where $\nu_a$ is the projectile-target relative velocity,
$\rho_b(E_b)=k_b\mu_b/[(2\pi)^3\hbar^2]$ is the density of the states
for the projectile $b$, $W_x$ is the imaginary part of the optical
potential describing $x+A$ elastic scattering, and $\psi^0_x(\vec{k}_b)$
is the relative state between $x$ and $A$, which governs the evolution
of $x$ after the collision, when particle $b$ is emitted with momentum $\vec{k}_b$
and the target remains in its ground state. This states satisfies the following
equation when representing on $x-A$ relative coordinates $\vec{r}_x$, where
the relevant coordinates are depicted in Fig.~\ref{fig:coordinates}
\begin{equation}
\label{eq:psi_x}
\langle \vec{r}_x|\psi^0_x(\vec{k}_b)\rangle = \int_0^\infty d\vec{r'_x}
G_x(\vec{r}_x,\vec{r'_x}) \langle \vec{r'_x}\chi_b(\vec{k}_b) | V_\mathrm{post}
| \Psi^{3b}\rangle, 
\end{equation}
where $G_x=1/(E_x-H_x)$ with the internal Hamiltonian
$H_x=T_x+U_x$ of $x-A$ subsystem and the relative energy $E_x$ between 
particles $x$ and $A$, $\chi_b$ is the distorted-wave describing
the scattering of $b$ in the final channel with respect to the $x-A$ subsystem,
, $V_\mathrm{post}=V_{bx}+U_{bA}-U_{b}$ (with $U_b$ the optical
potential in the outgoing channel) and $\Psi^{3b}$ is 
the three-body wave function, with boundary conditions corresponding to 
the incident $a$ particle.

Austern \textit{et al.} \cite{Austern87} suggested using the CDCC wave function
to approximate the three-body wave function, $\Psi^{3b}$, appearing in
Eq.(\ref{eq:psi_x}). Since the CDCC wave function is also a complicated
object which contains different partial wave components for the
$b-x$ subsystem, one needs to treat each partial wave equally.
In previous works\cite{Jin15,Jin17}, we have tested the validity
of $\ell=0$ case (deuterons and $^6$Li) and compared the calculation results with experimental
data. However, the IAV model has never been applied and tested for 
$\ell \geq 1$ cases to our knowledge. For that purpose, we employ the
distorted-wave Born approximation (DWBA), i.e., $\Psi^{3b}=\chi^{(+)}_a\phi_{a}$,
where $\chi_a^{(+)}$ is the distorted wave describing the $a+A$ elastic
scattering and $\phi_a$ is the projectile ground state wave function. Here
we will focus on $\ell=1$ case with $^7$Li.

Instead of using a three dimensional Jacobi basis, we expand the wave
function into partial wave eigenstates which depend on the magnitude of
the radius and angular momentum eigenstates. The orbital angular
momenta of three particles are coupled to total angular momentum $J$ and its third component,
for the incoming channels
\begin{equation}
|r_{bx}r_{a}\alpha_{in} \rangle = | r_{bx}r_{a} ((l_{a} (j_bj_x)s_{bx})J_{a}
(\lambda_a j_A)J_A) J M_J\rangle_{in},
\end{equation}
and for the outgoing channels
\begin{equation}
|r_{x}r_{b}\alpha_{out} \rangle = | r_{x}r_{b} ((l_x (j_xj_A)s_{xA})J_{x}
(\lambda_b j_b) J_b) JM_J\rangle_{out},
\end{equation}
where $j_b$, $j_x$ and $j_A$ are the internal spins of particles $b$, $x$,
and $A$ respectively and $l_{a}$, $\lambda_a$, $l_{x}$, and $\lambda_b$ are
the relative angular momentum of $b-x$, $a-A$, $x-A$, and $b-B^*$ respectively.

The angular momentum basis can be normalized as,
\begin{equation}
\langle r'_{bx}r'_{a}\alpha'_{in} |r_{bx}r_{a}\alpha_{in} \rangle
=\frac{\delta(r'_{bx}-r_{bx})}{r'_{bx}r_{bx}}\frac{\delta(r'_{a}-r_{a})}{r'_{a}r_{a}}
\delta_{\alpha'_{in},\alpha_{in}}, 
\end{equation}
and likewise for the outgoing basis.

In addition to that, a two body angular momentum basis for the $x-A$ subsystem
is used,
\begin{equation}
|r_x\beta\rangle = | r_x (l_xs_{xA})J_{x} M_{x} \rangle,
\end{equation}
therefore, the three body outgoing state can be decoupled by
\begin{equation}
|r_{x}r_{b}\alpha_{out} \rangle = \sum_{M_{{x}} M_{b}}
\langle  J_x M_{{x}}J_bM_{b}| JM_J \rangle |r_x\beta\rangle |r_bJ_bM_{b}\rangle,
\end{equation}
as well as the incoming state
\begin{equation}
|r_{bx}r_{a}\alpha_{in} \rangle = \sum_{M_{{a}} M_{A}}
\langle  J_a M_{{a}}J_AM_{A}| JM_J \rangle |r_{bx}J_aM_a\rangle |r_aJ_AM_{A}\rangle,
\end{equation}
where $M_{{x}}$, $M_{b}$, $M_a$, and $M_A$ are the third component
of $J_x$, $J_b$, $J_a$, and $J_A$ respectively.

By using the angular momentum basis defined above, we can rewrite Eq.(\ref{eq:psi_x})
as

\begin{equation}
\langle r_x\beta  | \psi^0_x(\vec{k}_b) \rangle = \int_0^{\infty}dr'_x r'^2_x G_x(r_x,r'_x,\beta)\rho(r'_x,\beta,\vec{k}_b),
\end{equation}
with
\begin{equation}
\label{eq:rho}
\rho(r'_x,\beta,\vec{k}_b) =
\langle  r'_x \beta \chi_b^{(-)}(\vec{k}_b) |
V_\mathrm{post}|\chi_a^{(+)}\phi_a \rangle .
\end{equation}

Since the incoming and outgoing channels are represented in their
natural set of Jacobi coordinate(see Fig.\ref{fig:coordinates}). 
A transformation from the sets $|r_{bx}r_{a}\alpha_{in}\rangle$
to $|r_xr_b\alpha_{out}\rangle$ is required. A partial wave
representation of this transformation is outlined in Ref.\cite{Balian1969} and can be written as an
integration over the cosine of the relative angle between $\vec{r}_x$ and $\vec{r}_b$.
All geometrical information is included in the coefficients
$\mathcal{G}^{out\gets in}_{\alpha_{in},\alpha_{out}}(r_xr_bx)$.
We give more details on these transformation in Appendix \ref{sec:appendix}. Additionally,
we only consider a central potential for $U_{bA}$. Then
inserting complete set of states in Eq. (\ref{eq:rho}) and making
use of the geometrical coefficients
$\mathcal{G}^{out\gets in}_{\alpha_{in},\alpha_{out}}(r_xr_bx)$, we arrive
at the following equation:
\begin{widetext}
\begin{equation}
\label{eq:rho}
\rho(r'_x,\beta,\vec{k}_b) = \sum_{\alpha_{out}}\int_0^\infty dr'_b r'^2_b
\Big\langle r'_x\beta \chi_b^{(-)}(\vec{k}_b) \Big| r'_xr'_b\alpha_{out}\Big\rangle
\sum_{\alpha_{in}} \int_{-1}^{1}dx V_\mathrm{post}(r'_xr'_bx\alpha_{out})
\mathcal{G}^{out\gets in}_{\alpha_{in},\alpha_{out}}(r'_xr'_bx)
\Big\langle r_{bx}r_a\alpha_{in}\Big|\chi_{a}^{(+)}\phi_a\Big\rangle ,
\end{equation}
\end{widetext}
with
\begin{widetext}
\begin{equation}
\Big\langle r'_x \beta \chi_b^{(-)}(\vec{k}_b) \Big| r'_xr'_b\alpha_{out}\Big\rangle
=\sum_{M_{{x}} M_{b}}
\langle  J_x M_{{x}}J_bM_{b}| JM_J \rangle \langle \chi_b^{(-)}(\vec{k}_b)  |r_bJ_bM_{b}\rangle
\delta_{\beta,J_x M_x} ,
\end{equation}
\end{widetext}
and
\begin{widetext}
\begin{equation}
\Big\langle r_{bx}r_a\alpha_{in}\Big|\chi_{a}^{(+)}\phi_a\Big\rangle
=\sum_{M'_{{a}} M_{A}}
\langle  J_a M'_{{a}}J_AM_{A}| JM_J \rangle
\langle r_{bx}J_aM'_a |\phi_a \rangle \langle r_aJ_AM_{A} | \chi_a^{(+)}\rangle ,
\end{equation}
\end{widetext}

The double differential cross section of NEB, which given by Eq.~(\ref{eq:neb})
can be represented with the angular momentum basis as
\begin{widetext}
\begin{equation}
\frac{d^2\sigma}{dE_bd\Omega_b}\Big|_{NEB} = -\frac{2}{\hbar \nu_a} \rho_b(E_b)
\sum_{\beta}\int_0^\infty dr_x r^2_x 
\big|\psi^{0}_x (r_x,\beta,\vec{k}_b)\big|^2 W_x(r_x,\beta).
\end{equation}
\end{widetext}


We consider the case of unpolarized beam, and aligned target. Moreover, 
we assume that the spin orientation of $b$ is not measured. In this 
situation, the cross section is obtained as an average of the initial angular momentum 
projections of $J_a$ and $J_A$, and a sum over the final projection of 
$J_b$. Thus
  
\begin{widetext}
\begin{equation}
\frac{d^2\sigma}{dE_bd\Omega_b}\Big|_{NEB} = -\frac{2}{\hbar \nu_a} \rho_b(E_b)\frac{1}{(2J_a+1)(2j_A+1)} 
\sum_{\beta}\sum_{M_am_Am_b}\int_0^\infty dr_x r^2_x 
\big|\psi^{0}_x (r_x,\beta,\vec{k}_b) \big|^2 W_x(r_x,\beta).
\end{equation}
\end{widetext}
where $m_A$ and $m_b$ are the third components of $j_A$ and
$j_b$, respectively

\section{Calculations}

\subsection{$^{58}$Ni($^7$Li,$\alpha$X)}
\begin{figure}[tb]
\begin{center}
 {\centering \resizebox*{0.9\columnwidth}{!}{\includegraphics{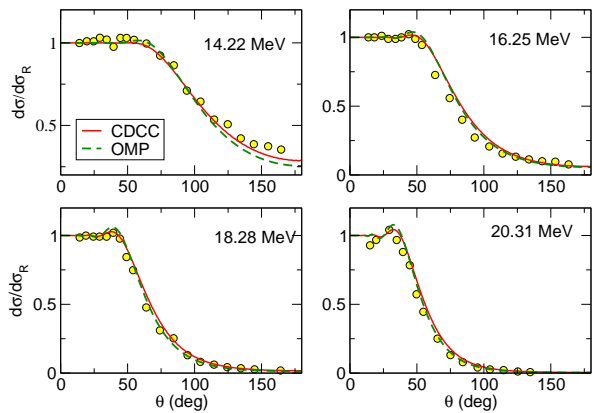}} \par}
\caption{\label{fig:elastic}(Color online)
Elastic scattering of $^7$Li + $^{58}$Ni at different incident energies.
The solid and dashed lines are the CDCC calculations and the optical model calculation
with the OMP of Cook \cite{Cook82}, respectively. Experimental data are taken from Ref. \cite{Pfeiffer73}.}
\end{center}
\end{figure}
To assess the validity of this partial wave expansion, we have done the
benchmark calculation comparing our earlier expansion given in Ref. \cite{Jin15}.
The numerical difference between these two method is less than $1\%$ by using 
the same input parameters. On the other hand, the well-known convergency 
problem in DWBA post form makes $\rho^{M_a}(r'_x,\beta)$ of Eq. (\ref{eq:rho}) 
long ranged. To overcome this issue, an identical prior form \cite{Jin15b} is used.

Now we present calculations for reactions induced
by a $^7$Li projectile and compare the calculated inclusive cross sections
with experimental data to assess the validity of the theory. In this case,
we compute the separate contributions for the elastic (EBU) and nonelastic (NEB) breakup cross sections.
For the former, we use the CDCC formalism, using the coupled-channels code FRESCO\cite{Thom88}.
This makes it possible to treat the EBU to all orders and should be
equivalent to the post-form three-body model of Austern et al.
For the NEB part, we use the DWBA version.

We consider the reaction $^{58}$Ni ($^7$Li,$\alpha X$) at energies around Coulomb barrier,
which allows us to compare with data from Ref.\cite{Pfeiffer73}.
The $^7$Li nucleus is treated in a two-cluster model $(\alpha + t)$.
Compared to the $(\alpha + d)$ two-cluster structure of $^6$Li, the main difference between
the two nuclei is the internal angular momentum $\ell$,
for $^6$Li $\ell = 0$, whereas for $^7$Li $\ell = 1$.
Furthermore the difference in the breakup threshold energy of
the two Li isotopes, $1.474$ MeV
for $\alpha + d$ breakup of $^6$Li compared to $2.468$ MeV for the
$\alpha + t$ breakup of $^7$Li is also important.

In order to test the validity of the $\alpha + t$ two cluster model for
$^7$Li, first the elastic scattering of the same reaction was studied
using the CDCC framework. The $\alpha + t$ interaction, which is
required to generate the $^7$Li ground state wave function as well as
the bound excited state and continuum wave functions, was taken from Ref.\cite{Buck88}.
This potential consists of a central and a spin-orbit component, of
Gaussian shape, with a fixed geometry and a parity-dependent depth.
The potential well depths were adjusted to give the correct binding
energy or resonance energy for bound or resonant states, respectively.
In order to achieve convergence of the calculated cross sections,
we needed to include $\alpha + t$ partial waves up to $\ell = 3$. For
the $f$ wave, a finer division of bins is used in order to reproduce
the $\ell = 3$ resonant states at 4.63 MeV ($7/2^{-}$)
and 6.68 MeV ($5/2^{-}$) correctly. The $^4$He-target interaction was obtained
from a Woods-Saxon potential fitted to the $12$ MeV $^4$He + $^{58}$Ni
elastic scattering data of Ref.\cite{Lee64} with the following parameters :
$V = 49.5$ MeV, $R_0 = 5.88$ fm, $a_0 = 0.5$ fm, $W = 11.0$ MeV,
$R_w = 5.69$ fm and $a_w = 0.5$ fm. The $^3$He-target interaction was
taken from the $8.95$ MeV $t$+$^{58}$Ni parameters of Ref.\cite{Fick84}.
For comparison, the optical model calculation using the potential of
Cook\cite{Cook82} was also performed. Fig. \ref{fig:elastic} shows the
elastic scattering of $^7$Li + $^{58}$Ni at different incident energies.
The data are taken from Ref.\cite{Pfeiffer73}.
The solid and dashed lines are, respectively, the CDCC and optical model
calculations. It can be seen that both the optical model
and CDCC calculations reproduce well the experimental data. This agreement
confirms the validity of the adopted $\alpha+$target and $t+$target optical potentials.

\begin{figure}[tb]
\begin{center}
 {\centering \resizebox*{0.9\columnwidth}{!}{\includegraphics{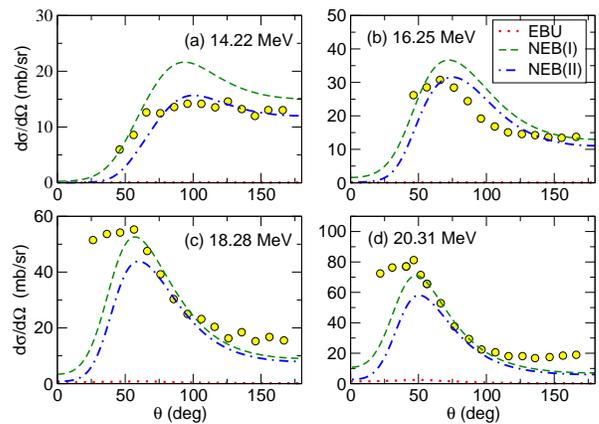}} \par}
\caption{\label{fig:inclusive}(Color online)
Angular distribution of $\alpha$ particles produced in the 
reaction $^7$Li + $^{58}$Ni at energies indicated by the labels.
The dotted, dashed and dot-dashed are, respectively, the EBU, NEB (DWBA) with $E_x<0$, and
NEB (DWBA) without $E_x<0$ components. The experimental data are taken from Ref.~\cite{Pfeiffer73}.
}
\end{center}
\end{figure}

Now the inclusive breakup cross section ($^7$Li,$\alpha X$) is discussed.
The EBU part was obtained from the CDCC calculation discussed above.
The NEB part was calculated with the IAV model using the DWBA
formalism without taking account the spin of particles.
There are two distinct contributions to the NEB cross sections, namely, 
that for $E_x>0$ case and that for $E_x<0$ case,
where $E_x$ is the final relative energy between $t$ and $^{58}$Ni.
For $E_x < 0$, this region would correspond to bound states
of the residual $^{61}$Cu system, that is, transfer.
The application of NEB formalism to transfer reactions is outlined in
Ref.\cite{Udagawa89} and recently applied to deuterons and $^6$Li induced
reactions\cite{Potel2017,Jin17}.
In Fig. \ref{fig:inclusive} the dotted, dashed and dot-dashed
lines are, respectively, the EBU (CDCC), NEB (DWBA) with $E_x<0$, and
NEB (DWBA) without $E_x<0$ components. First, it is noticeable that
the EBU part is negligible compared to the NEB component, which is in 
contrast to $^6$Li as reported in Ref.\cite{Jin17}. For the $^6$Li case, 
the contribution of EBU is small but non-negligible comparing to NEB. The difference
of these two nuclei will be discussed in the following section. 
Concerning the comparison of the calculations with experimental data,
we observe a good agreement with the data when including the $E_x<0$ part for
higher two energies and excluding the $E_x<0$ for lower two energies. 
The reason of that is not completely clear but it might be due to the 
fact that an energy-independent $t$+$^{58}$Ni potential has been 
employed, which will not describe correctly the low energy region 
(including the bound state part) of this system.  
A more relialistic description should be provided by a energy-dependent 
potential, extending also to negative energies. Such potentials were 
investigated in the past by Mahaux and Sartor \cite{Mahaux86} and are 
currently being revisited by several groups (see Ref. \cite{Dickhoff17} for a recent review).

\subsection{Comparison with the $^6$Li case}
\begin{figure}[tb]
\begin{center}
 {\centering \resizebox*{0.9\columnwidth}{!}{\includegraphics{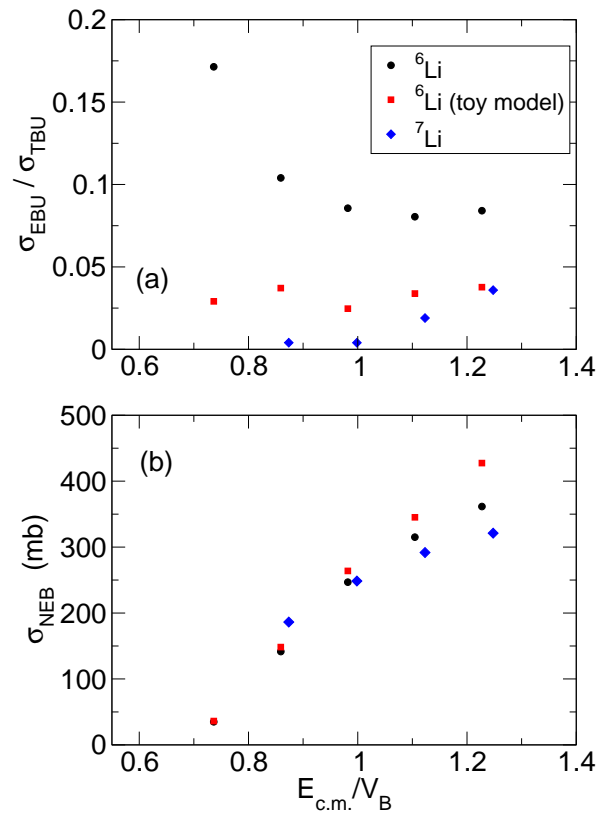}} \par}
\caption{\label{fig:ratios}(Color online)
(a) Ratios of  EBU over TBU (=EBU+NEB)
for $^{6,7}$Li + $^{58}$Ni systems. 
(b) NEB cross sections for $^{6,7}$Li + $^{58}$Ni systems. See text for details.}
\end{center}
\end{figure}

\begin{figure}[tb]
\begin{center}
 {\centering \resizebox*{0.9\columnwidth}{!}{\includegraphics{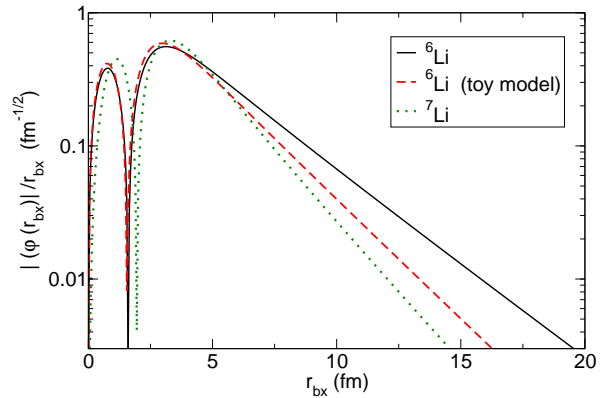}} \par}
\caption{\label{fig:wf}(Color online)
Projectile wave functions for $^{6,7}$Li. See text for details.}
\end{center}
\end{figure}

In this section, the difference between $^6$Li and $^7$Li on the $^{58}$Ni
target is discussed. The calculations of $^6$Li have been presented in Ref.{\cite{Jin17}}
In both cases, we have found that the NEB\footnote{NEB means the one with both $E_x>0$ and $E_x <0$ components} component
dominate the inclusive alphas. However it is interesting to compare the 
relative importance of EBU versus NEB on these two nuclei. 
In order to make a more meaningful comparison with 
these two nuclei, a toy model of $^6$Li is introduced by modifying the binding 
energy from $E_b=-1.474$ MeV to $E_b=-2.468$ MeV (that is, the 7Li binding energy).
Fig.\ref{fig:ratios} (a) plots the ration of the calculated EBU and TBU (=EBU+NEB) cross section as a function 
of the reduced energy $E_\mathrm{c.m.}/V_B$, with $V_B$ the energy of the 
Coulomb barrier, estimated as $V_B=Z_pZ_t/ [r_B(A^{1/3}_t + A^{1/3}_p)]$,
where $Z_p(Z_t)$ and $A_p(A_t)$ are the atomic number and atomic mass of 
projectile (target), respectively, and $r_B=1.44$ fm. The circles, 
squares and diamonds are respectively $^6$Li + $^{58}$Ni, $^6$Li$^\mathrm{toy}$ + $^{58}$Ni
and $^7$Li + $^{58}$Ni reaction systems. Several interesting features 
emerge from this plot: (i) First, for the lower binding energy, i.e., $^6$Li +$^{58}$Ni,
the elastic breakup component becomes more important as the energy decreases,
whereas for the energies above the Coulomb barrier, the ratio shows
an almost constant behavior; 
(ii) second, when increasing the binding of projectile, i.e., 
$^6$Li$^\mathrm{toy}$ + $^{58}$Ni, the 
elastic breakup component becomes comparively smaller; 
(iii) third, when changing the relative angular momentum in 
the projectile from $\ell=0$ to $\ell=1$, i.e., $^7$Li +$^{58}$Ni, the importance of elastic 
breakup component increases with the incident energy. These results
can be attributed to the fact that the EBU is a peripheral process 
and thereby highly sensitive to the tail of projectile wave function. 
In Fig.~\ref{fig:wf} , it can be clearly seen that $^6$Li has the 
longest tail among these three systems and this explains the larger EBU contribution. 
By contrast, since the wave function of $^7$Li is deeper hidden inside 
the Coulomb force, this case the $^7$Li projectile difficult to break in 
the relative low energies. 

Fig.\ref{fig:ratios} (b)  shows the NEB cross sections as a function of
the reduced energy $E_\mathrm{c.m.}/V_B$. It can be seen that the 
NEB cross section for these three systems are of similar magnitude. 
The NEB cross sections increase when changing the projectile binding energy 
by comparing with $^6$Li + $^{58}$Ni and $^6$Li$^\mathrm{toy}$ + $^{58}$Ni. 
However, the NEB cross section decrease when changing the internal relative
angular momentum from $\ell=0$ to $\ell=1$ ($^6$Li$^\mathrm{toy}$ to $^7$Li).
These behaviors indicate that the NEB is a volume process which comes from 
the interior part of projectile wave function and less sensitive to the 
internal structure of the projectile. This agrees with the fact that 
found in Ref. \cite{Moro2016}. 

\section{Summary and conclusions}

In summary, we addressed the problem of calculating the inclusive
breakup cross section for arbitrary $\ell$ values (with $\ell$ the orbital 
angular momentum between the clusters in the projectile ground state) within the closed-form DWBA 
model proposed in the 1980s by Ichimura, Austern, and Vincent\cite{IAV85}.
Moreover, numerical implementation of the model, more suitable 
for $\ell>0$ values, has been presented here.  

We have performed calculations for the $^{58}$Ni($^7$Li,$\alpha X$)
at energies around the Coulomb barrier. In this case, we find a good  
agreement between the experimental data and the IAV model. 

We also investigated effect of the internal structure of the projectile by 
comparing the $^7$Li inclusive breakup with $^6$Li. 
Although in both caes the $\alpha$ inclusive cross section is dominated by 
the NEB component, the EBU part is comparatively larger for the $^6$Li case. 
We interpret this as a consequence of the larger extension of the $^6$Li 
ground state wave function, due to its $\ell=0$ configuration.

The results presented in this work, along with those presented in previous 
works\cite{Jin15,Jin15b,Jin17}, indicate that the IAV model provides a 
reliable framework to calculate NEB cross sections. 
Possible applications to knockout reactions at intermediate energies
are currently under study.

\appendix
\section{Geometrical coefficient for coordinate transformation}
\label{sec:appendix}
In this section, we present the explicit expressions of the geometrical
coefficients $\mathcal{G}^{out\gets in}_{\alpha_{in},\alpha_{out}}(r'_xr'_bx)$.
These are given by
\begin{widetext}
\begin{eqnarray}
\label{eq:a1}
\mathcal{G}^{out\gets in}_{\alpha_{in},\alpha_{out}}(r'_xr'_bx)& =& \sum_{LS} (2S+1) \sqrt{(2J_{a}+1)(2J_{A}+1)(2J_{x}+1)(2J_{b}+1)}
\left\{ \begin{array}{ccc}   l_{x}  & s_{xA} & J_{x}  \cr
\lambda_{b} & j_{b}  & J_{b} \cr
L & S & J
\end{array}\right\} \,
\left\{   \begin{array}{ccc}   l_{a}  & s_{bx} & J_{a}  \cr
\lambda_{a} & j_{A}  & J_{A} \cr
L & S & J
\end{array}\right\}
 \nonumber \\
&& \times {8\pi^{2}} \, \sum_{M=-L}^{L}
\left\{ Y_{l_{x}}^{m_{l_{x}}*}(\hat r_{x}) \, Y^{m_{\lambda_{b}}*}_{\lambda_{b}}(\hat r_{b})   \right\}^{LM}
\left\{ Y^{m_{l_{a}}}_{l_{a}}(\reallywidehat {a \vec r_{x} - \vec r_{b} })
 \, Y^{m_{\lambda_{a}}}_{\lambda_{a}}( \reallywidehat {b\vec r_{x} + c \vec r_{b} } )   \right\}^{LM}
\nonumber \\
& & \times  (-)^{s_{bx}+2j_{A}+j_{x}+j_{b}} \sqrt{(2s_{xA}+1)(2s_{bx}+1)}
\left\{  \begin{array}{ccc}   j_{A}  & j_{x} & s_{xA}  \cr
j_{b} & S  & s_{bx} \cr
\end{array}\right\}
 \ \ .
\end{eqnarray}
\end{widetext}
The spherical harmonics $Y_l^m(\hat{r})$ depend on the angles $\hat{r}$
of the vector $\vec{r}$. For the evaluation, we choose $\vec{r}_{b}$ as
$z-$direction and $\vec{r}_x$ is in the $x-y$ plane:
\begin{equation}
\vec r_{b} = \left( \begin{array}{c}
0 \cr 0 \cr r_{b}
\end{array}\right)
\hspace{1cm} \vec r_{x} = \left( \begin{array}{c}
r_{x} \sqrt{1-x^{2}} \cr 0 \cr r_{x} x
\end{array}\right)  \ \ ,
\end{equation}
where $x$ is the cosine of the angle between $\vec{r}_b$ and $\vec{r}_x$.
In Eq.~(\ref{eq:a1}) the curly brackets grouping the spherical harmonics indicate that they
are coupled to a state of total orbital angular momentum $L$
and third component $M$. The mass ratios are given by
\begin{eqnarray}
a & =& \frac{m_{A}}{m_{A}+m_{x}} \\ \nonumber
b  & = & \frac{(m_{b}+m_{x}+m_{A})\, m_{x}}{(m_{A}+m_{x})(m_{b}+m_{x})} \\ \nonumber
c & =& \frac{m_{b}}{m_{b}+m_{x}} \ \ .
\end{eqnarray}
For this case, the coordinates of the incoming channel are given by
\begin{eqnarray}
r_{bx}(r_{x}r_{b}x) & =& \sqrt{a^{2} r_{x}^{2} +  r_{b}^{2} - 2 a  r_{x}r_{b}x}  \\ \nonumber
r_{a}(r_{x}r_{b}x) & =& \sqrt{b^2r_{x}^{2} + c^{2} r_{b}^{2} + 2 bc r_{x}r_{b}x}  \ \ .
\end{eqnarray}

\appendix*

\begin{acknowledgments}
The author is grateful to Antonio M.Moro and Filomena Nunes 
for a critical reading of the manuscript and helpful discussions.
This work has been supported by the National Science Foundation
under contract. No. NSF-PHY-1520972 with Ohio University.
\end{acknowledgments}

\bibliography{inclusive_prc.bib}

\begin{thebibliography}{28}%
\makeatletter
\providecommand \@ifxundefined [1]{%
 \@ifx{#1\undefined}
}%
\providecommand \@ifnum [1]{%
 \ifnum #1\expandafter \@firstoftwo
 \else \expandafter \@secondoftwo
 \fi
}%
\providecommand \@ifx [1]{%
 \ifx #1\expandafter \@firstoftwo
 \else \expandafter \@secondoftwo
 \fi
}%
\providecommand \natexlab [1]{#1}%
\providecommand \enquote  [1]{``#1''}%
\providecommand \bibnamefont  [1]{#1}%
\providecommand \bibfnamefont [1]{#1}%
\providecommand \citenamefont [1]{#1}%
\providecommand \href@noop [0]{\@secondoftwo}%
\providecommand \href [0]{\begingroup \@sanitize@url \@href}%
\providecommand \@href[1]{\@@startlink{#1}\@@href}%
\providecommand \@@href[1]{\endgroup#1\@@endlink}%
\providecommand \@sanitize@url [0]{\catcode `\\12\catcode `\$12\catcode
  `\&12\catcode `\#12\catcode `\^12\catcode `\_12\catcode `\%12\relax}%
\providecommand \@@startlink[1]{}%
\providecommand \@@endlink[0]{}%
\providecommand \url  [0]{\begingroup\@sanitize@url \@url }%
\providecommand \@url [1]{\endgroup\@href {#1}{\urlprefix }}%
\providecommand \urlprefix  [0]{URL }%
\providecommand \Eprint [0]{\href }%
\providecommand \doibase [0]{http://dx.doi.org/}%
\providecommand \selectlanguage [0]{\@gobble}%
\providecommand \bibinfo  [0]{\@secondoftwo}%
\providecommand \bibfield  [0]{\@secondoftwo}%
\providecommand \translation [1]{[#1]}%
\providecommand \BibitemOpen [0]{}%
\providecommand \bibitemStop [0]{}%
\providecommand \bibitemNoStop [0]{.\EOS\space}%
\providecommand \EOS [0]{\spacefactor3000\relax}%
\providecommand \BibitemShut  [1]{\csname bibitem#1\endcsname}%
\let\auto@bib@innerbib\@empty
\bibitem [{\citenamefont {Chattopadhyay}\ \emph {et~al.}(2016)\citenamefont
  {Chattopadhyay}, \citenamefont {Santra}, \citenamefont {Pal}, \citenamefont
  {Kundu}, \citenamefont {Ramachandran}, \citenamefont {Tripathi},
  \citenamefont {Sarkar}, \citenamefont {Sodaye}, \citenamefont {Nayak},
  \citenamefont {Saxena},\ and\ \citenamefont {Kailas}}]{Chattopadhyay16}%
  \BibitemOpen
  \bibfield  {author} {\bibinfo {author} {\bibfnamefont {D.}~\bibnamefont
  {Chattopadhyay}}, \bibinfo {author} {\bibfnamefont {S.}~\bibnamefont
  {Santra}}, \bibinfo {author} {\bibfnamefont {A.}~\bibnamefont {Pal}},
  \bibinfo {author} {\bibfnamefont {A.}~\bibnamefont {Kundu}}, \bibinfo
  {author} {\bibfnamefont {K.}~\bibnamefont {Ramachandran}}, \bibinfo {author}
  {\bibfnamefont {R.}~\bibnamefont {Tripathi}}, \bibinfo {author}
  {\bibfnamefont {D.}~\bibnamefont {Sarkar}}, \bibinfo {author} {\bibfnamefont
  {S.}~\bibnamefont {Sodaye}}, \bibinfo {author} {\bibfnamefont {B.~K.}\
  \bibnamefont {Nayak}}, \bibinfo {author} {\bibfnamefont {A.}~\bibnamefont
  {Saxena}}, \ and\ \bibinfo {author} {\bibfnamefont {S.}~\bibnamefont
  {Kailas}},\ }\href {\doibase 10.1103/PhysRevC.94.061602} {\bibfield
  {journal} {\bibinfo  {journal} {Phys. Rev. C}\ }\textbf {\bibinfo {volume}
  {94}},\ \bibinfo {pages} {061602} (\bibinfo {year} {2016})}\BibitemShut
  {NoStop}%
\bibitem [{\citenamefont {Pandit}\ \emph {et~al.}(2017)\citenamefont {Pandit},
  \citenamefont {Shrivastava}, \citenamefont {Mahata}, \citenamefont {Parkar},
  \citenamefont {Palit}, \citenamefont {Keeley}, \citenamefont {Rout},
  \citenamefont {Kumar}, \citenamefont {Ramachandran}, \citenamefont
  {Bhattacharyya}, \citenamefont {Nanal}, \citenamefont {Palshetkar},
  \citenamefont {Nag}, \citenamefont {Gupta}, \citenamefont {Biswas},
  \citenamefont {Saha}, \citenamefont {Sethi}, \citenamefont {Singh},
  \citenamefont {Chatterjee},\ and\ \citenamefont {Kailas}}]{Pandit17}%
  \BibitemOpen
  \bibfield  {author} {\bibinfo {author} {\bibfnamefont {S.~K.}\ \bibnamefont
  {Pandit}}, \bibinfo {author} {\bibfnamefont {A.}~\bibnamefont {Shrivastava}},
  \bibinfo {author} {\bibfnamefont {K.}~\bibnamefont {Mahata}}, \bibinfo
  {author} {\bibfnamefont {V.~V.}\ \bibnamefont {Parkar}}, \bibinfo {author}
  {\bibfnamefont {R.}~\bibnamefont {Palit}}, \bibinfo {author} {\bibfnamefont
  {N.}~\bibnamefont {Keeley}}, \bibinfo {author} {\bibfnamefont {P.~C.}\
  \bibnamefont {Rout}}, \bibinfo {author} {\bibfnamefont {A.}~\bibnamefont
  {Kumar}}, \bibinfo {author} {\bibfnamefont {K.}~\bibnamefont {Ramachandran}},
  \bibinfo {author} {\bibfnamefont {S.}~\bibnamefont {Bhattacharyya}}, \bibinfo
  {author} {\bibfnamefont {V.}~\bibnamefont {Nanal}}, \bibinfo {author}
  {\bibfnamefont {C.~S.}\ \bibnamefont {Palshetkar}}, \bibinfo {author}
  {\bibfnamefont {T.~N.}\ \bibnamefont {Nag}}, \bibinfo {author} {\bibfnamefont
  {S.}~\bibnamefont {Gupta}}, \bibinfo {author} {\bibfnamefont
  {S.}~\bibnamefont {Biswas}}, \bibinfo {author} {\bibfnamefont
  {S.}~\bibnamefont {Saha}}, \bibinfo {author} {\bibfnamefont {J.}~\bibnamefont
  {Sethi}}, \bibinfo {author} {\bibfnamefont {P.}~\bibnamefont {Singh}},
  \bibinfo {author} {\bibfnamefont {A.}~\bibnamefont {Chatterjee}}, \ and\
  \bibinfo {author} {\bibfnamefont {S.}~\bibnamefont {Kailas}},\ }\href
  {\doibase 10.1103/PhysRevC.96.044616} {\bibfield  {journal} {\bibinfo
  {journal} {Phys. Rev. C}\ }\textbf {\bibinfo {volume} {96}},\ \bibinfo
  {pages} {044616} (\bibinfo {year} {2017})}\BibitemShut {NoStop}%
\bibitem [{\citenamefont {Carnelli}\ \emph {et~al.}(2018)\citenamefont
  {Carnelli}, \citenamefont {Heimann}, \citenamefont {Pacheco}, \citenamefont
  {Arazi}, \citenamefont {Capurro}, \citenamefont {Niello}, \citenamefont
  {Cardona}, \citenamefont {de~Barbará}, \citenamefont {Figueira},
  \citenamefont {Hojman}, \citenamefont {Martí},\ and\ \citenamefont
  {Negri}}]{Carnelli18}%
  \BibitemOpen
  \bibfield  {author} {\bibinfo {author} {\bibfnamefont {P.}~\bibnamefont
  {Carnelli}}, \bibinfo {author} {\bibfnamefont {D.~M.}\ \bibnamefont
  {Heimann}}, \bibinfo {author} {\bibfnamefont {A.}~\bibnamefont {Pacheco}},
  \bibinfo {author} {\bibfnamefont {A.}~\bibnamefont {Arazi}}, \bibinfo
  {author} {\bibfnamefont {O.}~\bibnamefont {Capurro}}, \bibinfo {author}
  {\bibfnamefont {J.~F.}\ \bibnamefont {Niello}}, \bibinfo {author}
  {\bibfnamefont {M.}~\bibnamefont {Cardona}}, \bibinfo {author} {\bibfnamefont
  {E.}~\bibnamefont {de~Barbará}}, \bibinfo {author} {\bibfnamefont
  {J.}~\bibnamefont {Figueira}}, \bibinfo {author} {\bibfnamefont
  {D.}~\bibnamefont {Hojman}}, \bibinfo {author} {\bibfnamefont
  {G.}~\bibnamefont {Martí}}, \ and\ \bibinfo {author} {\bibfnamefont
  {A.}~\bibnamefont {Negri}},\ }\href {\doibase
  https://doi.org/10.1016/j.nuclphysa.2017.08.007} {\bibfield  {journal}
  {\bibinfo  {journal} {Nuclear Physics A}\ }\textbf {\bibinfo {volume}
  {969}},\ \bibinfo {pages} {94 } (\bibinfo {year} {2018})}\BibitemShut
  {NoStop}%
\bibitem [{\citenamefont {Sgouros}\ \emph {et~al.}(2016)\citenamefont
  {Sgouros}, \citenamefont {Pakou}, \citenamefont {Pierroutsakou},
  \citenamefont {Mazzocco}, \citenamefont {Acosta}, \citenamefont {Aslanoglou},
  \citenamefont {Betsou}, \citenamefont {Boiano}, \citenamefont {Boiano},
  \citenamefont {Carbone}, \citenamefont {Cavallaro}, \citenamefont {Grebosz},
  \citenamefont {Keeley}, \citenamefont {La~Commara}, \citenamefont {Manea},
  \citenamefont {Marquinez-Duran}, \citenamefont {Martel}, \citenamefont
  {Nicolis}, \citenamefont {Parascandolo}, \citenamefont {Rusek}, \citenamefont
  {S\'anchez-Ben\'{\i}tez}, \citenamefont {Signorini}, \citenamefont {Soramel},
  \citenamefont {Soukeras}, \citenamefont {Stefanini}, \citenamefont
  {Stiliaris}, \citenamefont {Strano}, \citenamefont {Strojek},\ and\
  \citenamefont {Torresi}}]{Sgouros16}%
  \BibitemOpen
  \bibfield  {author} {\bibinfo {author} {\bibfnamefont {O.}~\bibnamefont
  {Sgouros}}, \bibinfo {author} {\bibfnamefont {A.}~\bibnamefont {Pakou}},
  \bibinfo {author} {\bibfnamefont {D.}~\bibnamefont {Pierroutsakou}}, \bibinfo
  {author} {\bibfnamefont {M.}~\bibnamefont {Mazzocco}}, \bibinfo {author}
  {\bibfnamefont {L.}~\bibnamefont {Acosta}}, \bibinfo {author} {\bibfnamefont
  {X.}~\bibnamefont {Aslanoglou}}, \bibinfo {author} {\bibfnamefont
  {C.}~\bibnamefont {Betsou}}, \bibinfo {author} {\bibfnamefont
  {A.}~\bibnamefont {Boiano}}, \bibinfo {author} {\bibfnamefont
  {C.}~\bibnamefont {Boiano}}, \bibinfo {author} {\bibfnamefont
  {D.}~\bibnamefont {Carbone}}, \bibinfo {author} {\bibfnamefont
  {M.}~\bibnamefont {Cavallaro}}, \bibinfo {author} {\bibfnamefont
  {J.}~\bibnamefont {Grebosz}}, \bibinfo {author} {\bibfnamefont
  {N.}~\bibnamefont {Keeley}}, \bibinfo {author} {\bibfnamefont
  {M.}~\bibnamefont {La~Commara}}, \bibinfo {author} {\bibfnamefont
  {C.}~\bibnamefont {Manea}}, \bibinfo {author} {\bibfnamefont
  {G.}~\bibnamefont {Marquinez-Duran}}, \bibinfo {author} {\bibfnamefont
  {I.}~\bibnamefont {Martel}}, \bibinfo {author} {\bibfnamefont {N.~G.}\
  \bibnamefont {Nicolis}}, \bibinfo {author} {\bibfnamefont {C.}~\bibnamefont
  {Parascandolo}}, \bibinfo {author} {\bibfnamefont {K.}~\bibnamefont {Rusek}},
  \bibinfo {author} {\bibfnamefont {A.~M.}\ \bibnamefont
  {S\'anchez-Ben\'{\i}tez}}, \bibinfo {author} {\bibfnamefont {C.}~\bibnamefont
  {Signorini}}, \bibinfo {author} {\bibfnamefont {F.}~\bibnamefont {Soramel}},
  \bibinfo {author} {\bibfnamefont {V.}~\bibnamefont {Soukeras}}, \bibinfo
  {author} {\bibfnamefont {C.}~\bibnamefont {Stefanini}}, \bibinfo {author}
  {\bibfnamefont {E.}~\bibnamefont {Stiliaris}}, \bibinfo {author}
  {\bibfnamefont {E.}~\bibnamefont {Strano}}, \bibinfo {author} {\bibfnamefont
  {I.}~\bibnamefont {Strojek}}, \ and\ \bibinfo {author} {\bibfnamefont
  {D.}~\bibnamefont {Torresi}},\ }\href {\doibase 10.1103/PhysRevC.94.044623}
  {\bibfield  {journal} {\bibinfo  {journal} {Phys. Rev. C}\ }\textbf {\bibinfo
  {volume} {94}},\ \bibinfo {pages} {044623} (\bibinfo {year}
  {2016})}\BibitemShut {NoStop}%
\bibitem [{\citenamefont {Canto}\ \emph {et~al.}(2015)\citenamefont {Canto},
  \citenamefont {Gomes}, \citenamefont {Donangelo}, \citenamefont {Lubian},\
  and\ \citenamefont {Hussein}}]{Canto15}%
  \BibitemOpen
  \bibfield  {author} {\bibinfo {author} {\bibfnamefont {L.}~\bibnamefont
  {Canto}}, \bibinfo {author} {\bibfnamefont {P.}~\bibnamefont {Gomes}},
  \bibinfo {author} {\bibfnamefont {R.}~\bibnamefont {Donangelo}}, \bibinfo
  {author} {\bibfnamefont {J.}~\bibnamefont {Lubian}}, \ and\ \bibinfo {author}
  {\bibfnamefont {M.}~\bibnamefont {Hussein}},\ }\href {\doibase
  https://doi.org/10.1016/j.physrep.2015.08.001} {\bibfield  {journal}
  {\bibinfo  {journal} {Physics Reports}\ }\textbf {\bibinfo {volume} {596}},\
  \bibinfo {pages} {1 } (\bibinfo {year} {2015})},\ \bibinfo {note} {recent
  developments in fusion and direct reactions with weakly bound
  nuclei}\BibitemShut {NoStop}%
\bibitem [{\citenamefont {Ichimura}\ \emph {et~al.}(1985)\citenamefont
  {Ichimura}, \citenamefont {Austern},\ and\ \citenamefont {Vincent}}]{IAV85}%
  \BibitemOpen
  \bibfield  {author} {\bibinfo {author} {\bibfnamefont {M.}~\bibnamefont
  {Ichimura}}, \bibinfo {author} {\bibfnamefont {N.}~\bibnamefont {Austern}}, \
  and\ \bibinfo {author} {\bibfnamefont {C.~M.}\ \bibnamefont {Vincent}},\
  }\href {\doibase 10.1103/PhysRevC.32.431} {\bibfield  {journal} {\bibinfo
  {journal} {Phys. Rev. C}\ }\textbf {\bibinfo {volume} {32}},\ \bibinfo
  {pages} {431} (\bibinfo {year} {1985})}\BibitemShut {NoStop}%
\bibitem [{\citenamefont {Lei}\ and\ \citenamefont {{Moro}}(2015)}]{Jin15}%
  \BibitemOpen
  \bibfield  {author} {\bibinfo {author} {\bibfnamefont {J.}~\bibnamefont
  {Lei}}\ and\ \bibinfo {author} {\bibfnamefont {A.~M.}\ \bibnamefont
  {{Moro}}},\ }\href@noop {} {\bibfield  {journal} {\bibinfo  {journal} {Phys.
  Rev. C}\ }\textbf {\bibinfo {volume} {92}},\ \bibinfo {pages} {044616}
  (\bibinfo {year} {2015})}\BibitemShut {NoStop}%
\bibitem [{\citenamefont {Lei}\ and\ \citenamefont {Moro}(2015)}]{Jin15b}%
  \BibitemOpen
  \bibfield  {author} {\bibinfo {author} {\bibfnamefont {J.}~\bibnamefont
  {Lei}}\ and\ \bibinfo {author} {\bibfnamefont {A.~M.}\ \bibnamefont {Moro}},\
  }\href {\doibase 10.1103/PhysRevC.92.061602} {\bibfield  {journal} {\bibinfo
  {journal} {Phys. Rev. C}\ }\textbf {\bibinfo {volume} {92}},\ \bibinfo
  {pages} {061602} (\bibinfo {year} {2015})}\BibitemShut {NoStop}%
\bibitem [{\citenamefont {Potel}\ \emph {et~al.}(2015)\citenamefont {Potel},
  \citenamefont {Nunes},\ and\ \citenamefont {Thompson}}]{Pot15}%
  \BibitemOpen
  \bibfield  {author} {\bibinfo {author} {\bibfnamefont {G.}~\bibnamefont
  {Potel}}, \bibinfo {author} {\bibfnamefont {F.~M.}\ \bibnamefont {Nunes}}, \
  and\ \bibinfo {author} {\bibfnamefont {I.~J.}\ \bibnamefont {Thompson}},\
  }\href {\doibase 10.1103/PhysRevC.92.034611} {\bibfield  {journal} {\bibinfo
  {journal} {Phys. Rev. C}\ }\textbf {\bibinfo {volume} {92}},\ \bibinfo
  {pages} {034611} (\bibinfo {year} {2015})}\BibitemShut {NoStop}%
\bibitem [{\citenamefont {Potel}\ \emph
  {et~al.}(2017{\natexlab{a}})\citenamefont {Potel}, \citenamefont
  {Perdikakis}, \citenamefont {Carlson}, \citenamefont {Atkinson},
  \citenamefont {Dickhoff}, \citenamefont {Escher}, \citenamefont {Hussein},
  \citenamefont {Lei}, \citenamefont {Li}, \citenamefont {Macchiavelli},
  \citenamefont {Moro}, \citenamefont {Nunes}, \citenamefont {Pain},\ and\
  \citenamefont {Rotureau}}]{Pot17}%
  \BibitemOpen
  \bibfield  {author} {\bibinfo {author} {\bibfnamefont {G.}~\bibnamefont
  {Potel}}, \bibinfo {author} {\bibfnamefont {G.}~\bibnamefont {Perdikakis}},
  \bibinfo {author} {\bibfnamefont {B.~V.}\ \bibnamefont {Carlson}}, \bibinfo
  {author} {\bibfnamefont {M.~C.}\ \bibnamefont {Atkinson}}, \bibinfo {author}
  {\bibfnamefont {W.~H.}\ \bibnamefont {Dickhoff}}, \bibinfo {author}
  {\bibfnamefont {J.~E.}\ \bibnamefont {Escher}}, \bibinfo {author}
  {\bibfnamefont {M.~S.}\ \bibnamefont {Hussein}}, \bibinfo {author}
  {\bibfnamefont {J.}~\bibnamefont {Lei}}, \bibinfo {author} {\bibfnamefont
  {W.}~\bibnamefont {Li}}, \bibinfo {author} {\bibfnamefont {A.~O.}\
  \bibnamefont {Macchiavelli}}, \bibinfo {author} {\bibfnamefont {A.~M.}\
  \bibnamefont {Moro}}, \bibinfo {author} {\bibfnamefont {F.~M.}\ \bibnamefont
  {Nunes}}, \bibinfo {author} {\bibfnamefont {S.~D.}\ \bibnamefont {Pain}}, \
  and\ \bibinfo {author} {\bibfnamefont {J.}~\bibnamefont {Rotureau}},\ }\href
  {\doibase 10.1140/epja/i2017-12371-9} {\bibfield  {journal} {\bibinfo
  {journal} {The European Physical Journal A}\ }\textbf {\bibinfo {volume}
  {53}},\ \bibinfo {pages} {178} (\bibinfo {year}
  {2017}{\natexlab{a}})}\BibitemShut {NoStop}%
\bibitem [{\citenamefont {Carlson}\ \emph {et~al.}(2016)\citenamefont
  {Carlson}, \citenamefont {Capote},\ and\ \citenamefont {Sin}}]{Carlson2016}%
  \BibitemOpen
  \bibfield  {author} {\bibinfo {author} {\bibfnamefont {B.~V.}\ \bibnamefont
  {Carlson}}, \bibinfo {author} {\bibfnamefont {R.}~\bibnamefont {Capote}}, \
  and\ \bibinfo {author} {\bibfnamefont {M.}~\bibnamefont {Sin}},\ }\href
  {\doibase 10.1007/s00601-016-1054-8} {\bibfield  {journal} {\bibinfo
  {journal} {Few-Body Systems}\ }\textbf {\bibinfo {volume} {57}},\ \bibinfo
  {pages} {307} (\bibinfo {year} {2016})}\BibitemShut {NoStop}%
\bibitem [{\citenamefont {Moro}\ and\ \citenamefont {Lei}(2016)}]{Moro2016}%
  \BibitemOpen
  \bibfield  {author} {\bibinfo {author} {\bibfnamefont {A.~M.}\ \bibnamefont
  {Moro}}\ and\ \bibinfo {author} {\bibfnamefont {J.}~\bibnamefont {Lei}},\
  }\href {\doibase 10.1007/s00601-016-1085-1} {\bibfield  {journal} {\bibinfo
  {journal} {Few-Body Systems}\ }\textbf {\bibinfo {volume} {57}},\ \bibinfo
  {pages} {319} (\bibinfo {year} {2016})}\BibitemShut {NoStop}%
\bibitem [{\citenamefont {Lei}\ and\ \citenamefont {Moro}(2017)}]{Jin17}%
  \BibitemOpen
  \bibfield  {author} {\bibinfo {author} {\bibfnamefont {J.}~\bibnamefont
  {Lei}}\ and\ \bibinfo {author} {\bibfnamefont {A.~M.}\ \bibnamefont {Moro}},\
  }\href {\doibase 10.1103/PhysRevC.95.044605} {\bibfield  {journal} {\bibinfo
  {journal} {Phys. Rev. C}\ }\textbf {\bibinfo {volume} {95}},\ \bibinfo
  {pages} {044605} (\bibinfo {year} {2017})}\BibitemShut {NoStop}%
\bibitem [{\citenamefont {Luong}\ \emph {et~al.}(2013)\citenamefont {Luong},
  \citenamefont {Dasgupta}, \citenamefont {Hinde}, \citenamefont {du~Rietz},
  \citenamefont {Rafiei}, \citenamefont {Lin}, \citenamefont {Evers},\ and\
  \citenamefont {Diaz-Torres}}]{Luong13}%
  \BibitemOpen
  \bibfield  {author} {\bibinfo {author} {\bibfnamefont {D.~H.}\ \bibnamefont
  {Luong}}, \bibinfo {author} {\bibfnamefont {M.}~\bibnamefont {Dasgupta}},
  \bibinfo {author} {\bibfnamefont {D.~J.}\ \bibnamefont {Hinde}}, \bibinfo
  {author} {\bibfnamefont {R.}~\bibnamefont {du~Rietz}}, \bibinfo {author}
  {\bibfnamefont {R.}~\bibnamefont {Rafiei}}, \bibinfo {author} {\bibfnamefont
  {C.~J.}\ \bibnamefont {Lin}}, \bibinfo {author} {\bibfnamefont
  {M.}~\bibnamefont {Evers}}, \ and\ \bibinfo {author} {\bibfnamefont
  {A.}~\bibnamefont {Diaz-Torres}},\ }\href {\doibase
  10.1103/PhysRevC.88.034609} {\bibfield  {journal} {\bibinfo  {journal} {Phys.
  Rev. C}\ }\textbf {\bibinfo {volume} {88}},\ \bibinfo {pages} {034609}
  (\bibinfo {year} {2013})}\BibitemShut {NoStop}%
\bibitem [{\citenamefont {Pandit}\ \emph {et~al.}(2016)\citenamefont {Pandit},
  \citenamefont {Shrivastava}, \citenamefont {Mahata}, \citenamefont {Keeley},
  \citenamefont {Parkar}, \citenamefont {Rout}, \citenamefont {Ramachandran},
  \citenamefont {Martel}, \citenamefont {Palshetkar}, \citenamefont {Kumar},
  \citenamefont {Chatterjee},\ and\ \citenamefont {Kailas}}]{Pandit16}%
  \BibitemOpen
  \bibfield  {author} {\bibinfo {author} {\bibfnamefont {S.~K.}\ \bibnamefont
  {Pandit}}, \bibinfo {author} {\bibfnamefont {A.}~\bibnamefont {Shrivastava}},
  \bibinfo {author} {\bibfnamefont {K.}~\bibnamefont {Mahata}}, \bibinfo
  {author} {\bibfnamefont {N.}~\bibnamefont {Keeley}}, \bibinfo {author}
  {\bibfnamefont {V.~V.}\ \bibnamefont {Parkar}}, \bibinfo {author}
  {\bibfnamefont {P.~C.}\ \bibnamefont {Rout}}, \bibinfo {author}
  {\bibfnamefont {K.}~\bibnamefont {Ramachandran}}, \bibinfo {author}
  {\bibfnamefont {I.}~\bibnamefont {Martel}}, \bibinfo {author} {\bibfnamefont
  {C.~S.}\ \bibnamefont {Palshetkar}}, \bibinfo {author} {\bibfnamefont
  {A.}~\bibnamefont {Kumar}}, \bibinfo {author} {\bibfnamefont
  {A.}~\bibnamefont {Chatterjee}}, \ and\ \bibinfo {author} {\bibfnamefont
  {S.}~\bibnamefont {Kailas}},\ }\href {\doibase 10.1103/PhysRevC.93.061602}
  {\bibfield  {journal} {\bibinfo  {journal} {Phys. Rev. C}\ }\textbf {\bibinfo
  {volume} {93}},\ \bibinfo {pages} {061602} (\bibinfo {year}
  {2016})}\BibitemShut {NoStop}%
\bibitem [{\citenamefont {Shrivastava}\ \emph {et~al.}(2013)\citenamefont
  {Shrivastava}, \citenamefont {Navin}, \citenamefont {Diaz-Torres},
  \citenamefont {Nanal}, \citenamefont {Ramachandran}, \citenamefont {Rejmund},
  \citenamefont {Bhattacharyya}, \citenamefont {Chatterjee}, \citenamefont
  {Kailas}, \citenamefont {Lemasson}, \citenamefont {Palit}, \citenamefont
  {Parkar}, \citenamefont {Pillay}, \citenamefont {Rout},\ and\ \citenamefont
  {Sawant}}]{Shrivastava13}%
  \BibitemOpen
  \bibfield  {author} {\bibinfo {author} {\bibfnamefont {A.}~\bibnamefont
  {Shrivastava}}, \bibinfo {author} {\bibfnamefont {A.}~\bibnamefont {Navin}},
  \bibinfo {author} {\bibfnamefont {A.}~\bibnamefont {Diaz-Torres}}, \bibinfo
  {author} {\bibfnamefont {V.}~\bibnamefont {Nanal}}, \bibinfo {author}
  {\bibfnamefont {K.}~\bibnamefont {Ramachandran}}, \bibinfo {author}
  {\bibfnamefont {M.}~\bibnamefont {Rejmund}}, \bibinfo {author} {\bibfnamefont
  {S.}~\bibnamefont {Bhattacharyya}}, \bibinfo {author} {\bibfnamefont
  {A.}~\bibnamefont {Chatterjee}}, \bibinfo {author} {\bibfnamefont
  {S.}~\bibnamefont {Kailas}}, \bibinfo {author} {\bibfnamefont
  {A.}~\bibnamefont {Lemasson}}, \bibinfo {author} {\bibfnamefont
  {R.}~\bibnamefont {Palit}}, \bibinfo {author} {\bibfnamefont
  {V.}~\bibnamefont {Parkar}}, \bibinfo {author} {\bibfnamefont
  {R.}~\bibnamefont {Pillay}}, \bibinfo {author} {\bibfnamefont
  {P.}~\bibnamefont {Rout}}, \ and\ \bibinfo {author} {\bibfnamefont
  {Y.}~\bibnamefont {Sawant}},\ }\href {\doibase
  https://doi.org/10.1016/j.physletb.2012.11.064} {\bibfield  {journal}
  {\bibinfo  {journal} {Physics Letters B}\ }\textbf {\bibinfo {volume}
  {718}},\ \bibinfo {pages} {931 } (\bibinfo {year} {2013})}\BibitemShut
  {NoStop}%
\bibitem [{\citenamefont {Austern}\ \emph {et~al.}(1987)\citenamefont
  {Austern}, \citenamefont {Iseri}, \citenamefont {Kamimura}, \citenamefont
  {Kawai}, \citenamefont {Rawitscher},\ and\ \citenamefont
  {Yahiro}}]{Austern87}%
  \BibitemOpen
  \bibfield  {author} {\bibinfo {author} {\bibfnamefont {N.}~\bibnamefont
  {Austern}}, \bibinfo {author} {\bibfnamefont {Y.}~\bibnamefont {Iseri}},
  \bibinfo {author} {\bibfnamefont {M.}~\bibnamefont {Kamimura}}, \bibinfo
  {author} {\bibfnamefont {M.}~\bibnamefont {Kawai}}, \bibinfo {author}
  {\bibfnamefont {G.}~\bibnamefont {Rawitscher}}, \ and\ \bibinfo {author}
  {\bibfnamefont {M.}~\bibnamefont {Yahiro}},\ }\href {\doibase
  https://doi.org/10.1016/0370-1573(87)90094-9} {\bibfield  {journal} {\bibinfo
   {journal} {Physics Reports}\ }\textbf {\bibinfo {volume} {154}},\ \bibinfo
  {pages} {125 } (\bibinfo {year} {1987})}\BibitemShut {NoStop}%
\bibitem [{\citenamefont {Balian}\ and\ \citenamefont
  {Br{\'e}zin}(1969)}]{Balian1969}%
  \BibitemOpen
  \bibfield  {author} {\bibinfo {author} {\bibfnamefont {R.}~\bibnamefont
  {Balian}}\ and\ \bibinfo {author} {\bibfnamefont {E.}~\bibnamefont
  {Br{\'e}zin}},\ }\href {\doibase 10.1007/BF02710946} {\bibfield  {journal}
  {\bibinfo  {journal} {Il Nuovo Cimento B (1965-1970)}\ }\textbf {\bibinfo
  {volume} {61}},\ \bibinfo {pages} {403} (\bibinfo {year} {1969})}\BibitemShut
  {NoStop}%
\bibitem [{\citenamefont {Cook}(1982)}]{Cook82}%
  \BibitemOpen
  \bibfield  {author} {\bibinfo {author} {\bibfnamefont {J.}~\bibnamefont
  {Cook}},\ }\href {\doibase https://doi.org/10.1016/0375-9474(82)90513-9}
  {\bibfield  {journal} {\bibinfo  {journal} {Nuclear Physics A}\ }\textbf
  {\bibinfo {volume} {388}},\ \bibinfo {pages} {153 } (\bibinfo {year}
  {1982})}\BibitemShut {NoStop}%
\bibitem [{\citenamefont {Pfeiffer}\ \emph {et~al.}(1973)\citenamefont
  {Pfeiffer}, \citenamefont {Speth},\ and\ \citenamefont
  {Bethge}}]{Pfeiffer73}%
  \BibitemOpen
  \bibfield  {author} {\bibinfo {author} {\bibfnamefont {K.}~\bibnamefont
  {Pfeiffer}}, \bibinfo {author} {\bibfnamefont {E.}~\bibnamefont {Speth}}, \
  and\ \bibinfo {author} {\bibfnamefont {K.}~\bibnamefont {Bethge}},\ }\href
  {\doibase https://doi.org/10.1016/0375-9474(73)90084-5} {\bibfield  {journal}
  {\bibinfo  {journal} {Nuclear Physics A}\ }\textbf {\bibinfo {volume}
  {206}},\ \bibinfo {pages} {545 } (\bibinfo {year} {1973})}\BibitemShut
  {NoStop}%
\bibitem [{\citenamefont {Thompson}(1988)}]{Thom88}%
  \BibitemOpen
  \bibfield  {author} {\bibinfo {author} {\bibfnamefont {I.~J.}\ \bibnamefont
  {Thompson}},\ }\href@noop {} {\bibfield  {journal} {\bibinfo  {journal}
  {Comp. Phys. Rep.}\ }\textbf {\bibinfo {volume} {7}},\ \bibinfo {pages} {167
  } (\bibinfo {year} {1988})}\BibitemShut {NoStop}%
\bibitem [{\citenamefont {Buck}\ and\ \citenamefont {Merchant}(1988)}]{Buck88}%
  \BibitemOpen
  \bibfield  {author} {\bibinfo {author} {\bibfnamefont {B.}~\bibnamefont
  {Buck}}\ and\ \bibinfo {author} {\bibfnamefont {A.~C.}\ \bibnamefont
  {Merchant}},\ }\href {http://stacks.iop.org/0305-4616/14/i=10/a=002}
  {\bibfield  {journal} {\bibinfo  {journal} {Journal of Physics G: Nuclear
  Physics}\ }\textbf {\bibinfo {volume} {14}},\ \bibinfo {pages} {L211}
  (\bibinfo {year} {1988})}\BibitemShut {NoStop}%
\bibitem [{\citenamefont {Lee}\ and\ \citenamefont {Schiffer}(1964)}]{Lee64}%
  \BibitemOpen
  \bibfield  {author} {\bibinfo {author} {\bibfnamefont {L.~L.}\ \bibnamefont
  {Lee}}\ and\ \bibinfo {author} {\bibfnamefont {J.~P.}\ \bibnamefont
  {Schiffer}},\ }\href {\doibase 10.1103/PhysRev.134.B765} {\bibfield
  {journal} {\bibinfo  {journal} {Phys. Rev.}\ }\textbf {\bibinfo {volume}
  {134}},\ \bibinfo {pages} {B765} (\bibinfo {year} {1964})}\BibitemShut
  {NoStop}%
\bibitem [{\citenamefont {Fick}\ \emph {et~al.}(1984)\citenamefont {Fick},
  \citenamefont {Brown}, \citenamefont {Gr\"uebler}, \citenamefont
  {Hardekopf},\ and\ \citenamefont {Hanspal}}]{Fick84}%
  \BibitemOpen
  \bibfield  {author} {\bibinfo {author} {\bibfnamefont {D.}~\bibnamefont
  {Fick}}, \bibinfo {author} {\bibfnamefont {R.~E.}\ \bibnamefont {Brown}},
  \bibinfo {author} {\bibfnamefont {W.}~\bibnamefont {Gr\"uebler}}, \bibinfo
  {author} {\bibfnamefont {R.~A.}\ \bibnamefont {Hardekopf}}, \ and\ \bibinfo
  {author} {\bibfnamefont {J.~S.}\ \bibnamefont {Hanspal}},\ }\href {\doibase
  10.1103/PhysRevC.29.324} {\bibfield  {journal} {\bibinfo  {journal} {Phys.
  Rev. C}\ }\textbf {\bibinfo {volume} {29}},\ \bibinfo {pages} {324} (\bibinfo
  {year} {1984})}\BibitemShut {NoStop}%
\bibitem [{\citenamefont {Udagawa}\ \emph {et~al.}(1989)\citenamefont
  {Udagawa}, \citenamefont {Lee},\ and\ \citenamefont {Tamura}}]{Udagawa89}%
  \BibitemOpen
  \bibfield  {author} {\bibinfo {author} {\bibfnamefont {T.}~\bibnamefont
  {Udagawa}}, \bibinfo {author} {\bibfnamefont {Y.~J.}\ \bibnamefont {Lee}}, \
  and\ \bibinfo {author} {\bibfnamefont {T.}~\bibnamefont {Tamura}},\ }\href
  {\doibase 10.1103/PhysRevC.39.47} {\bibfield  {journal} {\bibinfo  {journal}
  {Phys. Rev. C}\ }\textbf {\bibinfo {volume} {39}},\ \bibinfo {pages} {47}
  (\bibinfo {year} {1989})}\BibitemShut {NoStop}%
\bibitem [{\citenamefont {Potel}\ \emph
  {et~al.}(2017{\natexlab{b}})\citenamefont {Potel}, \citenamefont
  {Perdikakis}, \citenamefont {Carlson}, \citenamefont {Atkinson},
  \citenamefont {Dickhoff}, \citenamefont {Escher}, \citenamefont {Hussein},
  \citenamefont {Lei}, \citenamefont {Li}, \citenamefont {Macchiavelli},
  \citenamefont {Moro}, \citenamefont {Nunes}, \citenamefont {Pain},\ and\
  \citenamefont {Rotureau}}]{Potel2017}%
  \BibitemOpen
  \bibfield  {author} {\bibinfo {author} {\bibfnamefont {G.}~\bibnamefont
  {Potel}}, \bibinfo {author} {\bibfnamefont {G.}~\bibnamefont {Perdikakis}},
  \bibinfo {author} {\bibfnamefont {B.~V.}\ \bibnamefont {Carlson}}, \bibinfo
  {author} {\bibfnamefont {M.~C.}\ \bibnamefont {Atkinson}}, \bibinfo {author}
  {\bibfnamefont {W.~H.}\ \bibnamefont {Dickhoff}}, \bibinfo {author}
  {\bibfnamefont {J.~E.}\ \bibnamefont {Escher}}, \bibinfo {author}
  {\bibfnamefont {M.~S.}\ \bibnamefont {Hussein}}, \bibinfo {author}
  {\bibfnamefont {J.}~\bibnamefont {Lei}}, \bibinfo {author} {\bibfnamefont
  {W.}~\bibnamefont {Li}}, \bibinfo {author} {\bibfnamefont {A.~O.}\
  \bibnamefont {Macchiavelli}}, \bibinfo {author} {\bibfnamefont {A.~M.}\
  \bibnamefont {Moro}}, \bibinfo {author} {\bibfnamefont {F.~M.}\ \bibnamefont
  {Nunes}}, \bibinfo {author} {\bibfnamefont {S.~D.}\ \bibnamefont {Pain}}, \
  and\ \bibinfo {author} {\bibfnamefont {J.}~\bibnamefont {Rotureau}},\ }\href
  {\doibase 10.1140/epja/i2017-12371-9} {\bibfield  {journal} {\bibinfo
  {journal} {The European Physical Journal A}\ }\textbf {\bibinfo {volume}
  {53}},\ \bibinfo {pages} {178} (\bibinfo {year}
  {2017}{\natexlab{b}})}\BibitemShut {NoStop}%
\bibitem [{\citenamefont {Mahaux}\ and\ \citenamefont
  {Sartor}(1986)}]{Mahaux86}%
  \BibitemOpen
  \bibfield  {author} {\bibinfo {author} {\bibfnamefont {C.}~\bibnamefont
  {Mahaux}}\ and\ \bibinfo {author} {\bibfnamefont {R.}~\bibnamefont
  {Sartor}},\ }\href {\doibase 10.1103/PhysRevLett.57.3015} {\bibfield
  {journal} {\bibinfo  {journal} {Phys. Rev. Lett.}\ }\textbf {\bibinfo
  {volume} {57}},\ \bibinfo {pages} {3015} (\bibinfo {year}
  {1986})}\BibitemShut {NoStop}%
\bibitem [{\citenamefont {Dickhoff}\ \emph {et~al.}(2017)\citenamefont
  {Dickhoff}, \citenamefont {Charity},\ and\ \citenamefont
  {Mahzoon}}]{Dickhoff17}%
  \BibitemOpen
  \bibfield  {author} {\bibinfo {author} {\bibfnamefont {W.~H.}\ \bibnamefont
  {Dickhoff}}, \bibinfo {author} {\bibfnamefont {R.~J.}\ \bibnamefont
  {Charity}}, \ and\ \bibinfo {author} {\bibfnamefont {M.~H.}\ \bibnamefont
  {Mahzoon}},\ }\href {http://stacks.iop.org/0954-3899/44/i=3/a=033001}
  {\bibfield  {journal} {\bibinfo  {journal} {Journal of Physics G: Nuclear and
  Particle Physics}\ }\textbf {\bibinfo {volume} {44}},\ \bibinfo {pages}
  {033001} (\bibinfo {year} {2017})}\BibitemShut {NoStop}%
\end{thebibliography}%
\end{document}